\documentclass[twocolumn]{aastex62}
\usepackage{amsfonts,amsmath,amssymb,amsthm,epsfig,graphicx,float,tabularx}
\usepackage{graphbox}
\usepackage{xcolor}
\usepackage{soul}

\shorttitle{Modeling Stellar Mergers and the Case of the B[e] Supergiant R4}
\shortauthors{Wu et al.}

\begin{document}

\title{The Art of Modeling Stellar Mergers and the Case of the B[e] Supergiant R4 in the Small Magellanic Cloud}

\correspondingauthor{S.C. Wu: scwu@caltech.edu}

\author{Samantha Wu}
\affiliation{Department of Astronomy \& Astrophysics, University of California, Santa Cruz, CA 95064, USA}
\affiliation{Niels Bohr Institute, University of Copenhagen, Blegdamsvej 17, 2100 Copenhagen, Denmark}
\affiliation{California Institute of Technology, 1200 East California Boulevard, MC 249-17, Pasadena, CA 91125, USA}

\author[0000-0001-5256-3620]{Rosa Wallace Everson}
\altaffiliation{NSF Graduate Research Fellow}
\affiliation{Department of Astronomy \& Astrophysics, University of California, Santa Cruz, CA 95064, USA}
\affiliation{Niels Bohr Institute, University of Copenhagen, Blegdamsvej 17, 2100 Copenhagen, Denmark}

\author[0000-0002-5965-1022]{Fabian R.~N. Schneider}
\affiliation{Zentrum f{\"u}r Astronomie der Universit{\"a}t Heidelberg, Astronomisches Rechen-Institut, M{\"o}nchhofstr.\ 12-14, 69120 Heidelberg, Germany}
\affiliation{Heidelberger Institut f{\"u}r Theoretische Studien, Schloss-Wolfsbrunnenweg 35, 69118 Heidelberg, Germany}

\author[0000-0002-8338-9677]{Philipp Podsiadlowski}
\affiliation{Department of Astronomy, Oxford University, Oxford OX1 3RH, UK}

\author[0000-0003-2558-3102]{Enrico Ramirez-Ruiz}
\affiliation{Department of Astronomy \& Astrophysics, University of California, Santa Cruz, CA 95064, USA}
\affiliation{Niels Bohr Institute, University of Copenhagen, Blegdamsvej 17, 2100 Copenhagen, Denmark}

\begin{abstract}
Most massive stars exchange mass with a companion, leading to evolution which is altered drastically from that expected of stars in isolation. 
Such systems result from unusual binary evolution pathways and can place stringent constraints on the physics of these interactions. 
We use the R4 binary system's B[e] supergiant, which has been postulated to be the product of a stellar merger, to guide our understanding of such outcomes by comparing observations of R4 to the results of simulating a merger with the 3d hydrodynamics code FLASH. 
Our approach tailors the simulation initial conditions to observed properties of R4 and implements realistic stellar profiles from the 1d stellar evolution code MESA onto the 3d grid, resolving the merger inspiral to within $0.02\, R_{\odot}$. We map the merger remnant into MESA to track its evolution on the HR diagram over a period of $10^4$ years.
This generates a model for a B[e] supergiant with stellar properties, age, and nebula structure in qualitative agreement with that of the R4 system.
Our calculations provide evidence to support the idea that R4's B[e] supergiant was originally a member of a triple system in which the inner binary merged after its most massive member evolved off the main sequence, producing a new object of similar mass but significantly more luminosity than the A supergiant companion. The code framework presented in this paper, which was constructed to model tidal encounters, can be used to generate accurate models of a wide variety of merger stellar remnants.
\end{abstract}

\section{Introduction}
Most massive stars exist in binaries or multiples, and the inevitable interaction with their companions via mass exchange dominates their evolution \citep{sana2012}. 
Of these interacting massive binaries, $\approx 25 \%$ will merge with their companion, which has significant implications for the resulting star's subsequent evolution \citep{podsiadlowski1992,sana2012,2014demink}.
These mergers and related binary interactions may give rise to peculiar phenomena such as gamma ray bursts \citep{2004ApJ...607L..17P,2004MNRAS.348.1215I,2011MNRAS.410.2458T}, luminous blue variables \citep{2014ApJ...796..121J}, and B[e] supergiants. In particular, \citet{podsiadlowski2006} argued that products of merger events are likely to be observed as B[e] supergiants as the merger adds mass to the core of the expanding primary star, modifying the core-envelope structure and altering the star's evolution so that it naturally populates the blue supergiant region of the HR diagram.

One such B[e] supergiant is observed in the R4 system in the Small Magellanic Cloud \citep{zickgraf1996} along with an A supergiant companion. The observed properties of this system exhibit an Algol-type paradox, which cannot be resolved by modeling the stars as evolving in isolation \citep{zickgraf1996,pasquali2000}. 
The B[e] supergiant in R4 thus appears to be an ideal candidate for a merged stellar remnant with clear observational constraints for the initial conditions  and end state of the system. However, very few such potential merger products have been identified from observations \citep[e.g.,][]{schneider2016}. 

Along with the rarity of observational constraints, realizing a fully self-consistent treatment of binary stellar mergers has been impeded by the complexity of the problem, which involves many physical processes spanning many orders of magnitude both spatially and temporally. One way to approach this is to divide the problem into separate phases, such that a different physical process dominates in each phase, and investigate each with a tailored numerical scheme \citep{podsiadlowski2001}. 

For example, when binary stars merge, the distorted internal structure of the stars has to be taken into consideration, and one must switch to a hydrodynamical description to follow the encounter. Hydrodynamical calculations need to be employed to study the deformations and exchange of energy and angular momentum, as well as the complete merger between the binary members \citep{sills1997,2013MNRAS.434.3497G,nandez2014,schneider2019}. 

After the dust has settled, one then has to update the stellar  models for the stars involved, and in the case of mergers one has to construct new models from scratch, often with highly unusual chemical compositions and physical conditions. The timescales for the stellar remnants to regain their thermal equilibrium are vastly longer than the timescales needed for dynamical equilibrium to be restored. In such cases, the merger remnant needs to be evolved in one dimension using an active stellar evolution code \citep{glebbeek2013,schneider2020}.

There is a history of over half a century of  stellar evolution calculations \citep[e.g.][]{1959ApJ...129..628H,bertelli1994,heger2000ApJ,meynetandmaeder2000,mesa2011}, and significant work on the hydrodynamics of stellar encounters has been done, in particular  in the context of smoothed particle hydrodynamics (SPH) simulations of blue stragglers  \citep[e.g.][]{freitag2005,dale2006,suzuki2007} and stellar collisions  \citep[e.g.][]{rasio1991,rasio1994,rasio1995,sills1997}. Pioneering work by \citet{sills1997}  emphasized the importance of bridging stellar evolution and SPH to achieve realistic collisional  products. 

The paucity of observations for possible mergers, let alone known merged remnants, motivates us to study the nature of unique systems such as R4 in order to be able to effectively constrain the physics of stellar mergers.
As a result, we choose to develop 3d hydrodynamical simulations of mergers using the R4 system as a guide (Section \ref{sec:initialconditions}). We select progenitor stars with structures that exhibit the desired core-envelope distinction and mass ratios that are consistent with the pre-merger system based on simple prescriptions for energy considerations (Section \ref{sec:initialmodels}).

Motivated by \citet{sills1997}, in this paper we  self-consistently implement MESA stellar profiles and corresponding equations of state onto our FLASH 3d grid simulation. 
In particular, we are able to resolve both the dense stellar core and the diffuse envelope on the grid with this realistic profile instead of appealing to the gravitational potential of a point mass to represent the core of the star (Section \ref{sec:simulation}), a distinction which is crucial to physically relevant simulations in the realm of stellar mergers and common envelope calculations.
This approach allows us to resolve the inspiral into the inner few solar radii of the star and enforce a physically motivated stopping criterion for the inspiral.
Finally, we map the merger remnant into a 1d stellar evolution code to track its position on the HR diagram as it regains thermal equilibrium. 
We compare the properties of the remnant and its surrounding nebula to observations of R4 in Sections \ref{sec:results} and \ref{sec:longtermevol}. In Sections \ref{sec:longtermevol} and \ref{sec:discussion}, we discuss how our methods, which encapsulate the merger process from inspiral to post-merger evolution, form a proof-of-concept for utilizing this setup to investigate similar systems. 

\section{Initial conditions}
\label{sec:initialconditions}
In this section, we determine which profiles are viable candidates for the pre-merger primary. We deduce minimum values of the mass unbound and energy injected into the remnant from observed properties of the R4 system. To determine which profiles can achieve these values, we look at a simple comparison of the binding energy of the envelope with the difference in initial and final orbital energies.  We also look at whether the energy expected to be injected into the remnant by the secondary during the merger is able to power the excess luminosity. This allows us to generate an initial grid of potential models that will be narrowed down further in Section \ref{sec:initialmodels}, using more careful considerations of the effects of drag on the dynamical inspiral phase of a merger.

\subsection{Observed properties of the R4 system}
\label{sec:observedproperties}
The R4 system as observed by \citet{zickgraf1996} consists of an evolved A supergiant and a B[e] supergiant companion separated by $a=23 $ AU. For the A supergiant, \citet{zickgraf1996} derive an effective temperature $T_{\rm eff}\approx 9500\text{--}11,000$K and fixed $\log g = 2.5$ from fitting ATLAS8 models.  In addition, they estimate mass of $12.9\, M_{\odot} \pm 2\, M_{\odot} $ from radial velocity (assuming $\mathbf{\sin^3{i}}=1)$. By iteratively fitting these parameters using the ATLAS8 models, \citet{zickgraf1996} find a radius of $R=33\, R_{\odot}$, which  gives a luminosity of $L \approx 10^4\, L_{\odot} $. They also derive a mass of $12.6\, M_{\odot}$ from the radius and $\log g$ values. Using a similar procedure, they find $T_{\rm eff}=27000 \, \rm{K}$, $\log g = 3.2$, $R=14\, R_{\odot}$, and $L=10^5\, L_{\odot}$ for the B[e] supergiant companion. The mass they derive from radial velocity (R-V) is $M=13.2\, M_{\odot} \pm 2\, M_{\odot}$, and from the radius and $\log g$  they find $M=11.3\, M_{\odot}$.

The effective temperature and luminosity of the B[e] star is well described by a supergiant with a ZAMS mass of $\approx 20\, M_{\odot}$, which is in stark contrast with the mass estimates from both radial velocity and $\log g$. This exemplifies the Algol-type paradox, where the B[e] star appears to have reached a very different stage in its evolution than the A supergiant despite their having similar measured masses.

The system exhibits a bipolar nebula with mean expansion velocity of $\sim 100$ km/s and an extension of $ \sim 2.4 $. pc \citep{pasquali2000}. Assuming a constant expansion velocity for the expanding material, the nebula's age can be estimated to be $\approx 10^4$ years.  
\citet{pasquali2000} conclude that the nebula was likely ejected from the B[e] supergiant as they find it to be nitrogen enriched as well as dynamically linked with the star.

\subsection{Evolutionary history of R4}
\label{sec:evolhist}
Given the observed separation,  it is reasonable to assume that the B[e] star and A supergiant companion have not interacted. Therefore, in what follows, we assume that the A supergiant has evolved independently as a single star. 

The observed effective temperature and luminosity of the B[e] component are not consistent with the evolution of a single star with the observationally derived mass estimate \citep{zickgraf1996}. In order to explain this tension, we may appeal to a process which is able to inject a significant amount of energy, resulting in higher luminosity. A stellar merger, in which the B[e] component was preceded by a close binary in a widely separated triple system with the A supergiant evolving independently, is one possibility. We refer to the more massive star in the close binary as the primary, and its less massive companion as the secondary. As a result of the merger, the secondary star injects energy, mass, and angular momentum  into the primary and unbinds a significant amount of envelope material. In this case, a merger  remnant might be left with properties similar to those  observed for the B[e] supergiant \citep{podsiadlowski2006}.

The existence and shape of the  nebula clearly  indicates that mass-loss occurred in a non-spherically symmetric fashion, which favors a dynamical event that occurred $\approx 10^4$ years ago. To constrain the initial conditions of this postulated merger event, we first assume that the system consisted previously of three stellar components born at the same time: star A, which evolved into the A supergiant; star B, which represents the aforementioned primary star in the merger that we postulate resulted in the observed B[e] supergiant; and star C, which represents the secondary star engulfed during the merger. Star A is likely to have evolved in isolation, so its age should help constrain the age of the R4 system. 

To estimate the age of star A, we run MESA \citep{mesa2011,mesa2013,mesa2015,mesa2018} simulations for stars evolving  into the supergiant phase with masses similar to those derived observationally. All models are generated with MESA version 10398. We use initial mass $12.5\, M_{\odot}$, which is within the range reported by \citet{zickgraf1996}. In all calculations, we start with pre-main-sequence models with an initial metallicity of $Z=0.1 Z_\sun$, given the system's location in the Small Magellanic Cloud. \footnote{For other parameters not listed, all MESA inlists are available upon request.}

\begin{figure}
    \includegraphics[width=0.94\columnwidth, align=t]{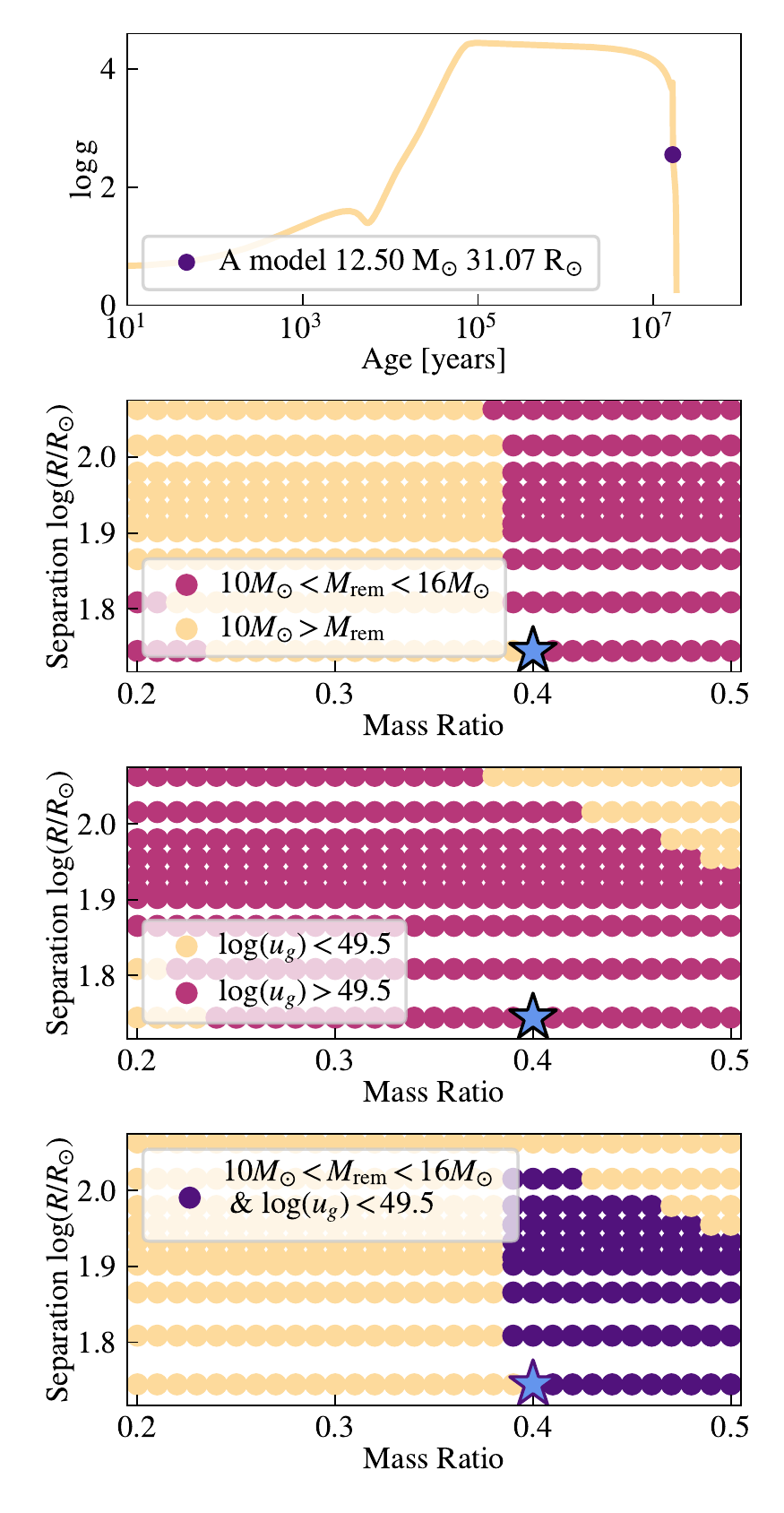}
    
    \caption{{\it Panel 1:} Evolution of $\log g$ over time for our A supergiant MESA model. The selected model (purple point) is within the mass, radius, and $\log g$ constraints of the observed R4 system. Model used for the hydrodynamical simulation performed with FLASH is shown as a blue star in each of the bottom three panels. {\it Panel 2:} Pairs of mass ratios $q$ and binary separations at the onset of common envelope, which we equate to the radius of a range of primary models, are shown as scatter points color-coded by the resultant remnant mass. Each scatter point represents a profile during the evolution of a MESA model with initial mass $13.0\, M_{\odot}$ and companion of mass ratio $0.2 \leq q \leq 0.5$. {\it Panel 3:} Scatter points represent the same pairs shown in Panel 2, color-coded by the energy in ergs released by the merger. {\it Panel 4:} Scatter points represent the same pairs as plotted in Panel 2. Purple dots are pairs that satisfy both criteria we seek, e.g. the following: $M_{\rm rem} = 13\, M_{\odot} \pm 3\, M_{\odot}$ and $\log(E[\rm{ergs}])>49.5$; peach points do not satisfy one or both of these criteria.} 
    \label{fig:qradiimfin}
\end{figure}

To select viable models for the A supergiant, we match the observed value of $\log g =2.5$ \citep{zickgraf1996} during the supergiant phase of evolution (Figure \ref{fig:qradiimfin}, top panel). This leads us to our model for the A supergiant, which has an age of $1.7\times10^7$ years with a mass of $12.5\, M_{\odot}$ and a radius of $31\, R_{\odot}$ at that age. The mass and radius successfully match the observed mass, radius, and $\log g$ values for the A supergiant.

Since the age of R4's nebula is of the order of $10^4$ years, the age of stars B and C at the time that the merger occurred must be approximately $10^4$ years less than the current age of star A. Dynamical mergers are driven by the expansion of the primary star. One possibility is that star B was crossing the Hertzsprung gap at that time, such that it was entering a slightly earlier stage of evolution than star A's current state (supergiant). For star B to have reached a similar stage of evolution as star A only $10^4$ years earlier means that it closely matched the evolution of star A. This suggests that the primary star in the merger had a slightly higher initial mass than that of the A supergiant. With this constraint in mind, we use the MESA code to generate models for star B, using the same inlist as for star A  but with an initial mass of $13\, M_{\odot}$. This choice is slightly arbitrary, but  similar masses ($<\pm1\, M_{\odot}$ variations) do not significantly affect the validity  of our conclusions. For consistency with the age of the nebula, we limit our consideration to models for the  primary which are $8\times 10^3$ years to a few $\times 10^4$ years younger than our A supergiant models. This restricts the size of the primary to  $55\, R_{\odot} \lesssim R \lesssim 120\, R_{\odot}$.

From the models within this range of radii, we select pre-merger primary profiles that have the capacity to release sufficient energy and unbind the required amount of mass. To estimate the radius at which energy will be released and mass unbound, we make use of the energy formalism, which equates the change in orbital energy of the secondary with the binding energy of the stellar envelope \citep{vandenheuvel1976,webbink1984,livio1988,iben1993}. We use the following form, calculated at each radial coordinate $r$:
\begin{equation}
    E_{\rm bind} (r) = \Delta E_{\rm orb} = - \frac{G M_1 M_2}{2R} + \frac{G M_{1,\rm enc} M_2}{2r}
\end{equation}
where $R$ and $M_1$ are the initial radius and mass of the primary, $M_2$ is the mass of the secondary, and $M_{1,\rm enc}$ is the enclosed mass of the primary at radius $r$. Here $E_{\rm bind} (r)$ is the binding energy of the stellar envelope beyond the chosen radial coordinate, and we use all available orbital energy to eject this portion of the envelope. Applying this formalism, we determine the coordinate in mass and radius where the change in orbital energy becomes larger than the binding energy of the envelope mass that is beyond this mass coordinate. 
We apply this criterion to a wide range of stellar profiles and mass ratios $q$, where $q M_{\rm B}=M_{\rm C}$ for primary mass $M_{\rm B}$ and companion mass $M_{\rm C}$. 

According to the energy formalism, the amount of orbital energy released at the mass coordinate of the crossing point is sufficient to unbind the envelope above this mass coordinate. As a result, the remaining mass of the primary star after the merger, $M_{\rm f}$, is equal to this mass coordinate. The mass of the remnant $M_{\rm rem}=M_{\rm f}+M_{\rm C}$ is shown for various combinations of radii and mass ratio in the second panel of Figure \ref{fig:qradiimfin}. We retain for further analysis the pairs of radii and mass ratio that produce  remnant masses of $13\, M_{\odot} \pm 3\, M_{\odot}$, within $2 \sigma$ of the approximate derived R-V mass for the  B[e] supergiant. In addition, the radius of each profile represents the pre-merger separation between the primary and its companion under the premise that the merger started as the companion came into contact with the remaining bound envelope. 

Moreover, the amount of orbital energy released at this mass coordinate provides an estimate of the amount of energy injected into the merger, which is expected to increase as the secondary plunges deeper into the core until it is tidally disrupted. 
At the end of the secondary's inspiral, $E_{\rm orb}/E_{\rm bind} \approx q^{-2/3}$, where $E_{\rm bind}$ is the binding energy. Since $q \lesssim 1$, the binding energy of the secondary, which will be deposited into the remnant is comparable or smaller than the orbital energy during the inspiral. The value of the orbital energy is therefore a simple proxy for how much energy will be deposited into the remnant. We select profiles with the capacity to inject more than $10^{49.5}$ ergs in addition to producing the desired remnant mass. The range of released energy $u_g$ for each pair of radius and mass ratio are shown in the third panel of Figure \ref{fig:qradiimfin}.
This estimate for the minimum injected energy was calculated assuming that the merger remnant needs to at least supply the current observed luminosity of $L_{\rm rem}\approx 10^5\, L_{\odot}$ for at least the age of the nebula, which is estimated to be $\approx 10^4$ years. 

The parameter space of potential primaries is represented by the intersection of the regions where $10\,  M_{\odot} < M_{\rm rem} < 16\,  M_{\odot}$ and where $\log(u_g)>49.5$. This intersection is shown in purple in the bottom panel of Figure \ref{fig:qradiimfin}. In the next section, we describe how we select our simulation initial conditions from this subset of viable pre-merger binaries.

\section{Methods}

\subsection{Initial Models}
\label{sec:initialmodels}
In this section, we select models to serve as the primary star in our hydrodynamical simulations. To narrow down the grid of models generated in Section \ref{sec:evolhist}, we focus our simulations on the dynamical inspiral phase of a merger and take into account the effects of drag during this phase. We decide on a model for the primary and mass ratio in which energy dissipation due to drag forces can unbind the necessary envelope mass so that the remnant mass $M_{\rm rem}$ matches mass estimates of R4's B[e] supergiant.

\begin{figure*}
    \includegraphics[width=\textwidth]{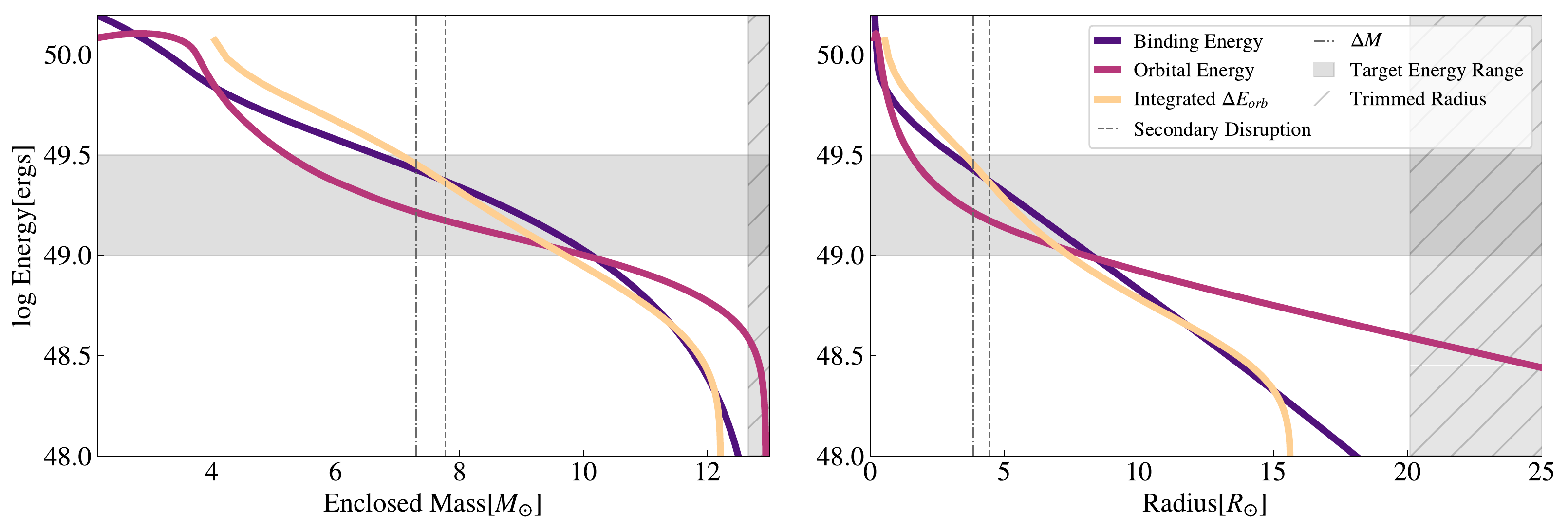}
    \caption{Relevant quantities for envelope unbinding during common envelope, shown on the left in mass coordinates and on the right in radius coordinates. These are presented for the model used as the initial condition for hydrodynamical simulation, with mass $12.9\, M_{\odot}$, radius $55\, R_{\odot}$, and secondary mass ratio $q=0.4$. The radius of the companion's disruption (dashed), total primary mass minus companion mass $\mathbf{\delta M}$ (dot-dashed), and released energy needed to match observed luminosity and age (grey region) are shown. The binding energy of material exterior to a given radial coordinate (purple), the corresponding difference in orbital energy relative to the initial orbit (magenta), and the integrated orbital energy dissipated from the inspiral (peach) are plotted against the mass and radius coordinate for each model.The region of the stellar profile removed for the FLASH simulations is shown in the grey hatched region.} 
    \label{fig:alphaplots}
\end{figure*}

In Figure \ref{fig:alphaplots}, we plot the 
properties of the stellar model which we have selected as the initial condition for the hydrodynamical simulation that we present in this paper. 
Figure \ref{fig:alphaplots} shows as a function of the radial (mass) coordinate the binding energy of the envelope, the change in the orbital energy from the start of the inspiral, and the energy dissipated by drag during the dynamical inspiral. 

We first note that the change in orbital energy curve (magenta) is above the binding energy curve (purple) at a relatively large outer radius. For radii beyond this crossing point, one can consider the envelope material, which contains a negligible fraction of the total mass,  to be easily ejected. This justifies our trimming of the stellar envelope at  $R\approx 20\, R_{\odot}$ when mapping into the  hydrodynamical simulations. The core and the envelope of the star can then be well-resolved in 3d without prohibitive computational costs. This is also motivated by \citet{podsiadlowski2001}, who note that the secondary's contact with the low-density outer envelope at the onset of mass transfer will produce a frictional luminosity  able to unbind stellar material well before the dynamical inspiral begins.

By trimming our envelope, we effectively focus our simulations on the dynamical inspiral phase and consider the envelope material beyond the crossing point to be ejected by the starting point of our simulations. Motivated by this, we consider modifications to the simple energy formalism used in section \ref{sec:evolhist} that take into account the importance of drag in driving the inspiral. We re-examine our profiles using Bondi-Hoyle-Lyttleton accretion (HLA) theory \citep{hoylelittleton1939} to calculate the energy dissipated due to drag, which is related to the gravitational drag force $F_{\rm d,HL}$ by
\begin{equation}
    \dot{E}_{\rm orb}(r) = -F_{\rm d,HL}v_{\infty}(r)
\end{equation}
where $v_{\infty}$ is the orbital speed of the secondary at a certain radius $r$, given by 
\begin{equation}
\label{eqn:vinfty}
    v_{\infty}=f_{\rm kep} v_{\rm kep}.
\end{equation}
Here $f_{\rm kep}$ is a factor describing the secondary's speed relative to the Keplerian velocity.
The gravitational drag force is
\begin{equation}
    F_{\rm d,HL}(r)=4\pi G^2 M_2^2 \rho_{\infty}(r)/v_{\infty}^2(r)
\end{equation}
where $M_2$ is the mass of the secondary and $\rho_{\infty}$ is the density of the primary at that radius. Using this formalism, we integrate $\dot{E}_{\rm orb}$ to find the total energy dissipated from the orbit $\Delta E_{\rm orb}$ along the inspiraling (non-circular) trajectory (peach curves in Figure \ref{fig:alphaplots}). We calculate the mass coordinate and energy where the curve for $\Delta E_{\rm orb}$ rises above the binding energy and take these values to be the mass unbound and energy released by the inspiral for that primary profile and given mass ratio $q$. 

We address these effects in more detail in Section \ref{sec:results} but note here that these values provide a reasonable lower limit to the energy injection, as the steep density gradients in the envelope would increase the energy dissipation rate  from the one described by $F_{\rm d, HL}$ \citep{macleod2017,de2020,everson2020}.

We also note that the dynamical inspiral will be terminated at an inner radius at which the secondary star would be tidally disrupted by the primary's core,
\begin{equation}
\label{eq:tidalradius}
    r_{\rm disrupt} = R_2\left(\frac{2\rho_{\rm enc}}{\rho_{\rm 2}} \right)^{1/3},
\end{equation}
where $\rho_{\rm enc}$ is the average enclosed density of the primary at $r_{\rm disrupt}$ and $R_2$ and $\rho_2$ are the radius and average density of the secondary  \citep{roche1849}.
The radius of disruption in Figure \ref{fig:alphaplots} shows the location where the secondary would begin to lose significant mass and can no longer be treated as a point mass as assumed by the equation of motion used to calculate the inspiral. In fact, we anticipate that at this radius the material of the secondary should begin to stream onto the core of the primary \citep{ivanova2002,ivanova2002PhDT}. 

In the binary model that we selected for the 3d  simulation (Figure~\ref{fig:alphaplots}), the HLA drag formalism predicts that  enough energy will be dissipated in order to unbind a mass comparable to the mass of the secondary. This is expected to occur at a similar but larger mass coordinate than that at which  the secondary would be disrupted by the primary's core, which was one of our key criteria. 
That is, the inspiral will likely terminate after the secondary dissipates enough energy to unbind the amount of mass needed to match the mass estimates of the B[e] progenitor. Our chosen model for the pre-merger system has a primary mass of $12.9\,  M_{\odot}$, secondary mass ratio $q=0.4$, and radius of $55\,  R_{\odot}$. Its age is $\approx 10^4$ years younger than the A supergiant model described in Section \ref{sec:initialconditions}. Since we avoid prohibitively high resolution in our 3d hydrodynamics simulation by using only the inner $20 R_{\odot}$ of the primary profile and representing the secondary with a point mass (Section \ref{sec:simulation}), our simulation setup has the ability to most closely reproduce an initial condition with smaller primary and secondary sizes. This consideration guided us to select this model, which pairs the smallest allowed values of separation and mass ratio as predicted by the overlap region shown in the bottom panel of Figure \ref{fig:qradiimfin}. Note that the methods of this section yield other valid pre-merger models that satisfy these conditions and are within the overlap region of Figure \ref{fig:qradiimfin}; in this paper, we present the results of simulating one of these options. 

\subsection{Description of simulation}
\label{sec:simulation}
We map the density, pressure, temperature, and composition of the 1d MESA profile onto a 3d grid using FLASH \citep{2000ApJS..131..273F} version 4.3, a grid-based adaptive mesh refinement hydrodynamics code.  Our setup is adapted from \citet{2013ApJ...767...25G}, but it uses an extended Helmholtz equation of state \citep{2000ApJS..126..501T}  instead of a polytropic EOS. In addition, we track the composition of elements as described in \citet{lawsmith2019}.

In order to resolve the inspiral near the core while utilizing a reasonable amount of computational resources, we trim the profile to $20\, R_{\odot}$ for the simulation, which we justify with analytical  results presented in Figure \ref{fig:alphaplots}. The computational domain is cubical with volume $(80\,  R_{\odot})^3$ and is initially composed of an $8^3$ block grid with a minimum cell size of $0.019\,  R_{\odot}$. 

To setup the initial condition, we initially relax the stellar profile for a few  dynamical times. During relaxation, a point mass (constructed to represent the secondary) is placed at $15 R_{\odot}$, initially at rest. The velocity of the secondary is then gradually increased during the relaxation process until it attains an approximate  Keplerian velocity as determined by the enclosed mass at $15\,  R_{\odot}$. The  mass of the secondary is $5.18\,  M_{\odot}$, corresponding to $q=0.4$. Once relaxation ends, the primary model is in hydrostatic equilibrium and the inspiral trajectory is calculated self-consistently. The properties of the merger outcome are found to be rather insensitive to the exact initial conditions of the secondary's velocity, provided it is close to Keplerian. This assumption is justified by the inspiral calculations presented in Section \ref{sec:initialmodels}. We stop the simulation once the particle reaches the tidal radius (Equation \ref{eq:tidalradius}). We compare the numerical trajectories with the analytic/HLA drag predictions presented in Section~\ref{sec:initialmodels} and find that while both show a dynamical plunge, the  secondary in the hydrodynamical simulation inspirals  at a slightly faster rate. This is expected to be the case as the HLA drag coefficients are systematically  lower than those derived when the stellar density gradients are included, as shown by \citet{2015ApJ...803...41M} and \citet{2017ApJ...838...56M}.

\begin{figure*}
    \centering
    \includegraphics[width=\textwidth]{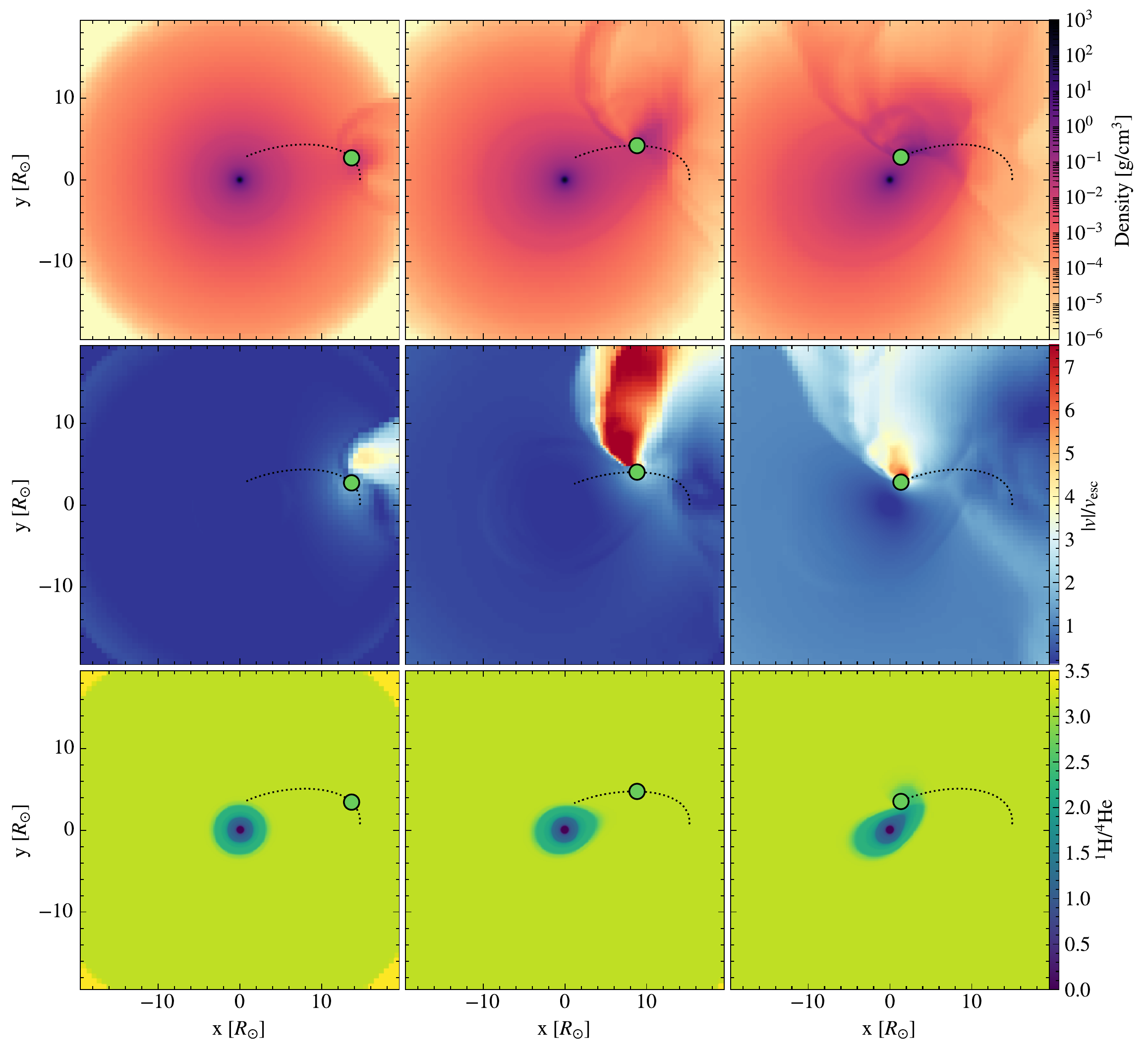}
    \caption{2D slices in the $xy$ plane, showing snapshots of the simulation at $t=14\, t_{\rm dyn}, t=24\, t_{\rm dyn},$ and $t=t_{\rm disrupt}=28t_{\rm dyn}$, where $t_{\rm dyn}$ is the core dynamical time and $t_{\rm disrupt}$ is the time when the secondary reaches the tidal disruption radius. Increasing time is read from left to right. {\it Top:} Density of all material. {\it Middle:} Velocity magnitude divided by the core escape velocity $v_{\rm  esc}\approx 10^8 \, \rm{cm/s}$. {\it Bottom:} Ratio of $^1\rm{H}$ and $^4\rm{He}$ mass fractions. The secondary is shown as a green dot and its inspiral is shown by the dotted line. The color of the star is chosen from the colorbar in the bottom panel, based on the secondary's hydrogen-to-helium ratio.}
    \label{fig:xyslices} 
\end{figure*}

\subsection{Constructing MESA Models for the  Remnant}
\label{subsec:MESArelax}

To understand the merger remnant in terms of observables, we map our simulation results into MESA and allow the resulting profiles to evolve further. 
Applying the relaxation module to the merger model, we relax the composition, then the entropy, using MESA's \texttt{inlist\_massive\_defaults} along with an inlist specifying parameters for relaxation.

Before importing the 3d simulation results to MESA, the material of the bound primary mass has to be combined with the secondary.
At the end of the hydrodynamical simulation, the secondary has reached a radius at which it would tidally disrupt due to the gravitational influence of the primary's core ($r_{\rm disrupt}$), causing material and energy to be deposited  around that radius. 
We approximate the tidal disruption of the secondary by adding the mass and chemical composition of the secondary to the bound primary material in the vicinity of the tidal disruption radius. Using a MESA model of a $ 5.18 M_{\odot}$ main sequence star to determine the secondary's chemical and thermal profile, we distribute the mass of the secondary across the outer mass shells of the bound primary material such that the greatest amount of mass is added around the mass coordinate of $r_{\rm disrupt}$, with the remaining mass added in a tail skewing towards larger radius.
This in turn determines the distribution of the combined chemical and thermal structure. We then sort the shells of the combined remnant profile by entropy, such that entropy increases with radius \citep{lombardi2002}.

We map this remnant into the 1d stellar evolution code MESA. This entails making a MESA model of a star whose total mass is equal to the sum of the bound primary mass and the secondary mass, as well as having chemical and thermal structure that matches that of the combined merger remnant. Using the methods outlined in \citet{schneider2016,schneider2019, schneider2020}, we achieve a 1d MESA model with a structure that is a close match to that of the combined merger remnant described above.

To account for the deposition of energy from the disruption of the secondary, we add heat to the remnant during the MESA evolution. A total energy equal to the binding energy of the secondary is injected into the remnant during evolution at shells in the vicinity of $r_{\rm disrupt}$. This is certainly a lower limit to the amount of energy injected into the remnant, as we must also consider the secondary's remaining orbital energy. However, it is not clear what proportion of the remaining orbital energy is dissipated into the remnant rather than being used to spin off the  envelope of the primary once the secondary is tidally disrupted. A detailed understanding of this   requires 3d hydrodynamical simulations of this stage that resolve both objects in order to determine the resultant energy dissipation and rotational profiles. For simplicity, here we take the conservative approach of only adding the binding energy. Each shell receives the same heat per unit mass at a constant rate $\approx E_{\rm bind} \times 10^{-7} \rm{s}^{-1}$ until energy equal to the binding energy has accumulated, at $ \approx 6$ years. This is much shorter than the total time over which the remnant is evolved ($\gtrsim 10^5$ years).

We evolve the resulting relaxed combined model in MESA using  \texttt{inlist\_massive\_defaults} along with a base inlist for evolution until the end of helium burning.

\section{Hydrodynamical Simulation}
\label{sec:results}

\begin{figure*}
    \centering
    \hspace*{-1.0em}\includegraphics[width=0.95\textwidth]{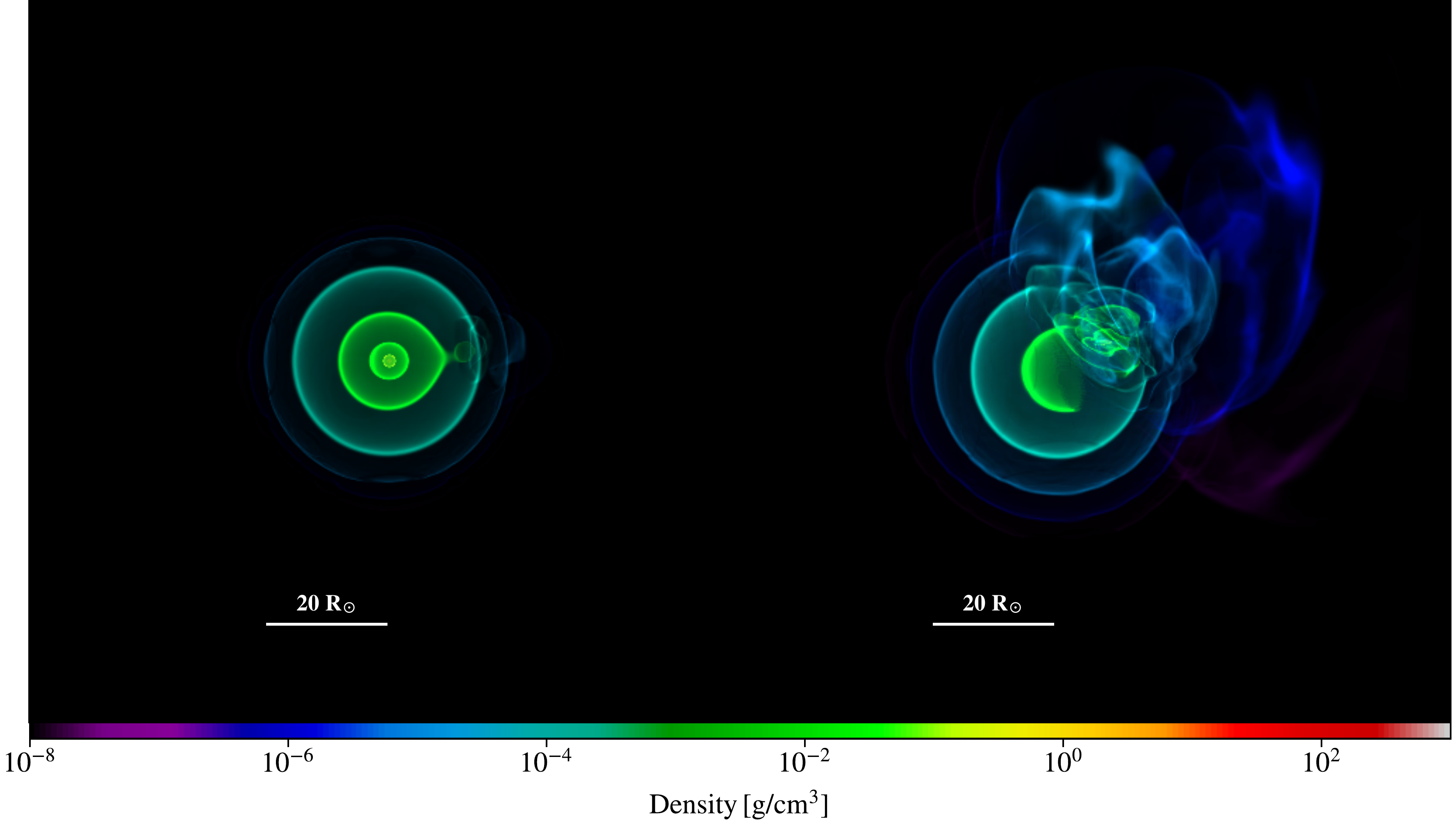}
    \caption{ {\it Left Panel:} 3d rendering of density of all material in the 3d hydrodynamical simulation, shown at $t = 14 t_{\rm dyn}$. The diameter of material depicted is $40 R_{\odot}$ across. {\it Right Panel:} 3d rendering of density of only unbound material at $t=t_{\rm disrupt}$.  }
    \label{fig:3drenderdens} 
\end{figure*}

In this section we present the results of our FLASH simulation modeling the  merger of a binary chosen in Section \ref{sec:initialmodels} to represent the progenitor of R4's B[e] supergiant. In our simulation, the initial mass of the primary model is $13\, M_{\odot}$ and has a companion-to-primary mass ratio of $q=0.4$, corresponding to a secondary with mass $5.18\, M_{\odot}$. The primary star's initial radius before trimming is $55\,  R_{\odot}$; after being trimmed to $20\,  R_{\odot}$, the pre-merger primary mass is $12.7\,  M_{\odot}$. In addition to the simulation presented here, we ran simulations with different initial conditions that also met the requirements outlined in Sections \ref{sec:initialconditions} and \ref{sec:initialmodels}. We find the results presented here to be representative, as only minor differences are observed.

\subsection{Dynamical Inspiral}
\label{sec:dynamicalinspiral}
As the inspiral progresses over time (left to right in the top 3 panels of Figure \ref{fig:xyslices}), the secondary rapidly plunges into the core of the primary via dynamical inspiral. We expected this steep plunge-in from our initial conditions, as we placed the secondary deep in the envelope of the primary where the inspiral would be driven by strong drag forces.

In Section \ref{sec:initialmodels} we narrowed down our profiles using  HLA drag theory to predict the amount of unbound mass and released energy, but the results of such an approach are thought to serve as a rough estimate for  these values. In practice, the initial conditions of the simulation push the limits of the power that HLA drag theory possesses to predict the path of our expected inspiral, since HLA is predicated on the assumption that the inspiral deviates only mildly  from a Keplerian orbit throughout. In a steep spiral-in the trajectory is far from Keplerian, as we see in the progression of the inspiral for the $55\,  R_{\odot}$ profile in Figure \ref{fig:xyslices}. 

However, based on the ideas of \citet{macleod2017}, the steep density gradient of the primary's envelope and the high $q$ value indicate that the effects of drag can be approximated by multiplying the drag force $F_{\rm d, HL}$ by a constant coefficient $C_d$, applied only in the tangential direction and opposing the direction of motion. To guide our understanding of how these factors steepened the inspiral, we calculate an average $C_d$ by comparing the timescale of the inspiral with the ratio of the change in orbital energy, $\Delta E$, and the rate of energy dissipation by gravitational drag, $\dot{E}$. We use the following relation
\begin{equation}
    C_d=\Delta E_{\rm orb}/(F_{\rm d,HL}v_2t_{\rm orb})
\end{equation}
with the average values of density and velocity for $r_{\rm disrupt} < R < 20\, R_{\odot}$ and the change in orbital energy from $R=20\, R_{\odot}$ to $R=r_{\rm disrupt}$, and we find an average $C_d=2.6$. Here $r_{\rm disrupt}$ is the tidal disruption radius as in Equation \ref{eq:tidalradius} (see section \ref{sec:endofinspiral} for value). Thus on average, the drag force is a factor of $2.6$ higher than the HLA prediction, which is in agreement with the results of \citet{macleod2017}. A higher drag force implies that we would expect the orbital energy to be dissipated at a smaller mass coordinate and the inspiral to proceed more rapidly than the one predicted by HLA. This aligns with the results of our simulation, which tends to unbind slightly more mass and has a steeper inspiral trajectory than that predicted in Section \ref{sec:initialmodels}. In addition, the change in orbital energy deviates from that commonly assumed by the $\alpha$ formalism, which assumes circular orbits. As Figure \ref{fig:alphaplots}  shows, the change in orbital energy due to drag dissipation (peach) rises above the binding energy curve (purple) at different mass coordinates than the difference in orbital energy calculated under the assumptions of the $\alpha$ formalism (magenta).

\subsection{End of Inspiral}
\label{sec:endofinspiral}
The simulation is evolved until the point mass representing the secondary reaches the disruption radius, at $3.85 \, R_{\odot}$. In the bottom three panels of Figure \ref{fig:xyslices}, we see that as the inspiral proceeds (left to right), the core of the primary becomes distorted and even partially disrupted once the secondary reaches its own tidal disruption radius. 
At this stage, $\approx 4.6\, M_{\odot}$ of mass is unbound.
Our calculations of the initial conditions predicted that the secondary would unbind $\approx 5\,  M_{\odot}$ by the time the engulfed star reached its tidal disruption radius for both primaries, which agrees well with the total amount of mass found to be ejected in our simulation. We also note that $ \lesssim 8 \%$ of the original primary mass or $\lesssim 1\, M_{\odot}$ has left the simulation box over the duration of the simulation.  

\subsection{Remnant and Nebula}
Once the secondary has reached the tidal disruption radius, we treat the merger remnant as composed of material from the disrupted secondary and the bound mass of the primary. At this point in the simulation, $8.3\,  M_{\odot}$ of primary material remains bound. The bottom right panel of Figure \ref{fig:xyslices} shows the ratio of $^1\rm{H}$ to $^4\rm{He}$ mass fractions at the end of the simulation for both the primary and secondary. The composition of the bound remnant will be mixed in the outer layers with the different composition of the secondary.

The nebula produced by the merger will consist of unbound material whose velocity is greater than the escape velocity of the core of the primary star. In the middle three panels of Figure \ref{fig:xyslices}, we plot the velocity magnitude divided by the escape velocity of the core throughout the simulation. As the inspiral progresses (left to right), more material reaches large enough velocities to be able to escape. The plunge-in of the secondary up to the tidal disruption radius highly disturbs the envelope material and causes an asymmetric ejection of unbound material; the 3d renderings in Figure \ref{fig:3drenderdens} depict how the primary star's envelope is affected at $t = 14 t_{\rm dyn}$ and $t = 28 t_{\rm dyn}$. Although the material in the path of the inspiral is preferentially accelerated along the path of least resistance, a significant fraction of material at a radial distance $\gtrsim 5 R_{\odot}$ becomes unbound in all directions by the time the secondary reaches the tidal disruption radius. 

The total kinetic energy of the unbound material is $3.2\times10^{50}$ ergs and its average velocity is $1.7\times10^8$ cm/s, which is $1.8\,  v_{\rm esc}$ (the core's escape velocity). As shown in the center-right panel of Figure \ref{fig:xyslices}, the majority of the unbound material moves initially at speeds that are in excess of those observed in R4's nebula, which exhibits velocities of $\approx 10^7$ cm/s. As the ejected nebula material expands, it will sweep up the surrounding material and, as a result, it will decelerate. The displaced volume  as derived  from the size of the observed nebula implies that the ejected material will sweep a mass that is larger than its  own ($\approx 4-5\,  M_{\odot}$) and  thus is expected to decelerate significantly.

The morphology of the unbound material in the simulations once the secondary has reached the disruption radius provides us with a qualitative picture for the shape of the nebula resulting from the merger. The 3d rendering in the right panel of Figure \ref{fig:3drenderdens} of the density of unbound material forms an asymmetric bipolar structure. \citet{pasquali2000} conclude from kinematics that R4's nebula also is not strictly bipolar. However, R4's nebula clearly has a complicated structure and resolving its  morphology requires higher-resolution observations. In addition, any detailed comparison of the merger ejecta with simulations will need long-term modelling of the ejecta's expansion including  interactions with the ISM and the stellar winds, and the illumination  from the merger remnant.  

\begin{figure}[h]
    \includegraphics[width=\columnwidth]{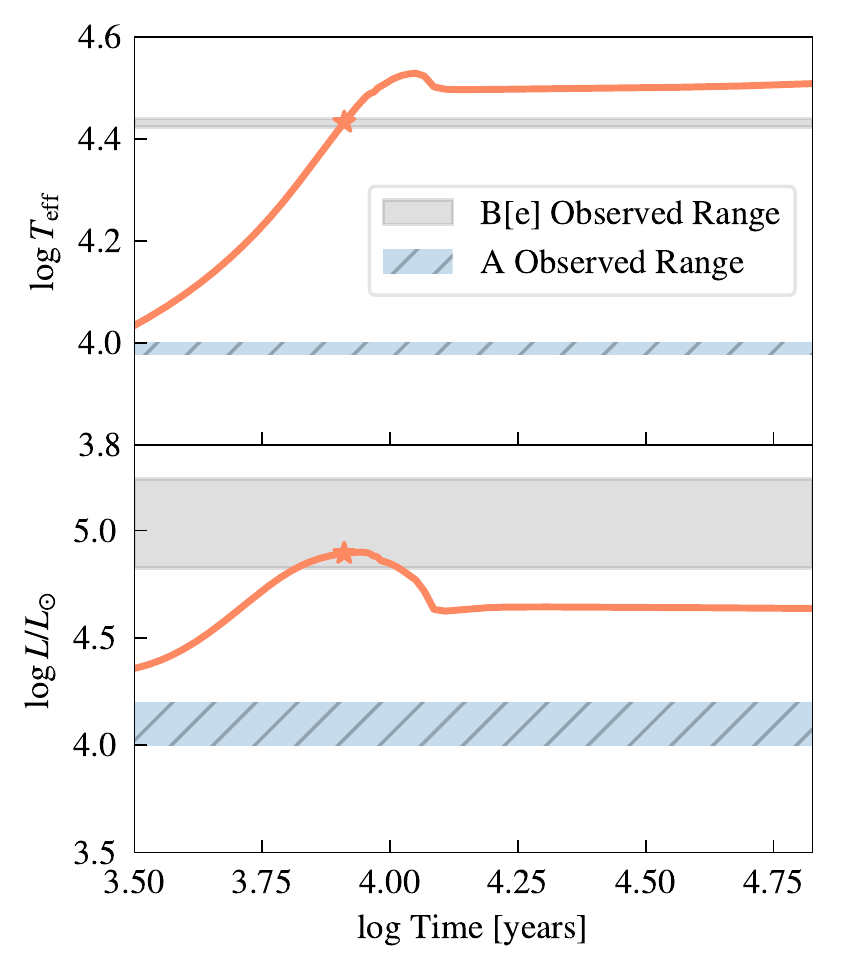}
    \caption{Evolution of the temperature and luminosity of the merger remnant in MESA. Here time is measured in years since the merger. 
    The grey bar shows the range of $T_{\rm eff}$ and $L$ observed for the B[e] component. The remnant remains hotter and more luminous than the observed A component (blue hatched bar) throughout the cooling period, the duration of which is also in agreement with the age of the remnant as derived by the age of the nebula. The star symbol denotes the model for the merger remnant described in Section 
    \ref{sec:longtermevol} that best exhibits the observed properties.}
    \label{fig:mesaevol}
\end{figure}

\section{Long Term Evolution}
\label{sec:longtermevol}

Figure \ref{fig:mesaevol} shows the track of the remnant's evolution in effective temperature and luminosity over time. \citet{zickgraf1996} determined the effective temperature and luminosity of the B[e] star by taking their best fits to the effective temperature $T_{\rm eff}$ and $\log g$ values, then calculating the bolometric luminosity using the radius they found from $\log g$ and their spectroscopic mass with $\sin i \approx 1$. From Figure 8 of \citet{zickgraf1996}, we deduce approximate error bars of $T_{\rm eff} = 27000\,  \rm{K} \pm 500\,  \rm{K}$ and $\log g = 3.2 \pm 0.175$ (mean values correspond to those cited in Section \ref{sec:observedproperties} for the B[e] star). We also deduce a bolometric luminosity $L=10^{4.95}\, L_{\odot}$ from Figure 10 of \cite{zickgraf1996} and derive error bars on the luminosity measurement from those on $T_{\rm eff}$ and $\log g$. The $1 \sigma$ ranges for $T_{\rm eff}$ and $L$ are shown in grey in Figure \ref{fig:mesaevol}.

We cite values for the evolution of the merger remnant from the $55\,  R_{\odot}$ simulation here. The remnant attains $T_{\rm eff}=27000\,  \rm{K}$ at 
$8.14\times 10^3$ years (Figure \ref{fig:mesaevol}, top panel). This model (star symbol in Figure \ref{fig:mesaevol}) has $\log g=3.34$, corresponding to a mass of $12.9\,  M_{\odot}$ and radius $12.73\,  R_{\odot}$. 

The $T_{\rm eff}$, $\log g$, and mass values are within the errors for the observed values, and the radii resulting from these values are close to the radius $14\,  R_{\odot}$ derived from the observed values of \citet{zickgraf1996}. The luminosity is $L=10^{4.9}\,  L_{\odot}$ (Figure \ref{fig:mesaevol}, bottom panel), again very near the derived value of \cite{zickgraf1996}. 

Our long-term evolution of the merger remnant produces a model which achieves the same effective temperature, luminosity, radius, and mass as the observed B[e] supergiant. This model exhibits all the observed characteristics at $8\times10^3$ years post-merger. We compare in Figure \ref{fig:mesaevol} the evolution of the merger remnant to the observed properties of star A (blue hatched region), which has a similar mass but has solely undergone single-star  evolution.  The evolution of the merger remnant starts to deviate from the evolution of star A soon after the merger, as large amounts of heat are injected deep into its interior that must be radiated away. This  allows the merger remnant to remain extremely luminous for a cooling phase of about $10^5$ years. 

Since in this scenario we expect the nebula to be the result of ejected material from the merger, we take the age of the remnant to be equal to the age of the nebula, which is derived from the observed expansion velocity and nebula size to be $\sim 10^4$ years. Thus our model is able to successfully reproduce the observed properties of the B[e] supergiant at the expected age of the remnant. Our evolved merger remnant therefore constitutes a viable model for the B[e] supergiant of the R4 system.

It is important to note that the late-time evolution ($t \gtrsim 10^5$ years) of our merger remnant is sensitive to our mixing prescription and whether we include rotation. Details of how the merger remnant may evolve on the HR diagram after the cooling period will be explored in future work.

Ultimately, achieving our goal of studying the long-term evolution of the remnant depended on our ability to map our merger remnant from the 3d hydrodynamical code FLASH into MESA, a 1d stellar evolution code. Bridging this gap allowed us to make more concrete statements about how applicable our merger models truly are to a particular system. Furthermore, we were able to treat the long-term evolution as a natural continuation of the merger process for the system by mapping the final conditions of the 3d simulation onto the initial conditions of the 1d simulation. The combination of our highly resolved hydrodynamical simulations with the stellar evolution code allowed us to investigate various stages of the merger that proceed on widely different timescales, all of which are needed in order to accurately compute the evolution of systems hosting multiple interacting stars.

\section{Summary and Conclusions}
\label{sec:discussion}

We have studied the evolutionary history of the R4 system using 3d hydrodynamical simulations and a 1d stellar evolution code to model its B[e] supergiant. We chose this system because it has been postulated that a binary stellar merger produced the B[e] supergiant. Furthermore, the R4 system was especially conducive to the study of binary stellar mergers since the observations provided enough constraints on the properties of the system to develop sensible initial conditions (Section \ref{sec:initialconditions}). Observations of the nebula size and expansion velocity limited the age of the nebula, which is a proxy for the time since merger. We also appealed to the large observed separation between the stars in the R4 system to deduce that the A supergiant companion evolved independently, and to the observed luminosity of the B[e] star to set a lower limit on the amount of energy injected into the merger.

Using initial conditions driven by the observed properties of the R4 system, we have simulated a binary stellar merger using a 3d hydrodynamics code and mapped the merger remnant into a 1d stellar evolution code to study its long-term evolution. As a result, we were able to compare the R4 system to the remnant at a time since merger that matches the nebula age. We find that our methods produce a model for the merger remnant at the appropriate time whose stellar properties are in good agreement with the B[e] supergiant. The long-term evolution also suggests that the remnant is still undergoing a cooling phase after the merger, during which period it remains extremely luminous and attains the paradoxically high effective temperature and luminosity of the B[e] supergiant.

Even with the observational constraints, some degeneracy remains in the choice of progenitor masses and separations (Section \ref{sec:evolhist}). We have chosen to simulate a particular combination that satisfies the initial conditions outlined in Section \ref{sec:initialconditions}. The success of the exercise applied to this choice serves as a proof-of-concept for the methods laid out in this paper to study similar problems by transitioning between FLASH and MESA. In particular, the dynamical inspiral of the merger process was consistently extended to the long-term evolution of the remnant. The process may be applied to different progenitors and different systems to generate models of a variety of merger remnants, which, as thoughtfully argued by \citet{sana2012}, are expected to be common.

Note that the MESA models for the merger remnant were evolved without rotation. During the plunging of the companion, a significant fraction of the orbital angular momentum is transferred to the unbound envelope material in our simulations. At the time the companion reaches the tidal disruption radius, it has a sub-Keplerian velocity $v\approx 0.6\, v_{\rm kep}$.  The companion will be disrupted at this stage and 
its orbital angular momentum is expected to be effectively transferred to the merger remnant. Assuming that the secondary's angular momentum is distributed uniformly over the remnant, we can calculate the remnant's final rotation velocity.
The angular velocity that the remnant gains from merging with the secondary is given by $\Delta \Omega=\frac{\Delta J}{I}$, where  $I=\frac{2}{5}(M_{\rm bound}+M_2)R^2$ is the moment of inertia of the remnant. Here $\Delta J = f_{\rm kep}M_2\sqrt{G M_{\rm bound} R}$ is the additional angular momentum of the secondary, where the orbital speed of the secondary is measured relative to the Keplerian velocity as in Equation \ref{eqn:vinfty}. Evaluating this at the tidal disruption radius we find $f_{\rm kep}=0.6$, which implies that the addition of the angular momentum of the secondary is expected to spin up the merger remnant to $\approx 36 \%$ of its breakup velocity. 
In our parameter space of initial conditions, there are some initial conditions that would give the final merger product even higher rotation as the final ratio of $M_2$ to $M_{\rm bound}$ could be closer to unity.
Although in principle this rotation would serve as another reservoir of energy for the remnant to draw upon, more detailed study of the angular momentum transport throughout the remnant is required to robustly estimate its dissipation rate. Here we take the simplest approach of not including rotation in our MESA model, which will provide a lower limit to the luminosity of the merger product over its thermal timescale.  
 
In addition to a more careful treatment of rotation in our remnant, we envision many other avenues for extending our work in the future. It would be useful to investigate the details of how late-term evolution of the merger remnant, after the thermal relaxation period is over, will proceed. In particular, the effects of different mixing prescriptions and of the ensuing rotational profile of the remnant ought to be better quantified. Furthermore, while in this work the secondary was modeled as a point mass, endeavors to model both primary and secondary using realistic stellar profiles from MESA are already underway. This would allow the 3d hydrodynamical simulation to realistically follow the inspiral all the way to merger instead of stopping at the secondary's tidal disruption radius. A simulation using realistic profiles would moreover resolve how the material of the secondary streams on to the core of the primary. This would provide a more accurate model for the size and shape of the merger remnant and would also narrow the uncertainty in the mixing prescription used to map the remnant into MESA.

To conclude, the proposed numerical formalism may be applied to model the outcomes of mergers, collisions, and tidal disruptions \citep{lawsmith2019,lawsmith2020}.
On the timescale of the study, we could only hope to explore in detail merely some subset of the interesting possible encounters that could have given rise to the R4 system (Figure \ref{fig:qradiimfin}). In the near future, we hope to develop a comprehensive model database of remnants and their predicted observational outcomes for a range of events.
Such a formalism would serve as a valuable theoretical counterpart to the increasing number of merger remnant products expected to be uncovered in future observational surveys \citep{sana2013,almeida2017,mahy2020}.

\acknowledgements{We gratefully acknowledge helpful discussions with D. Lee, S. de Mink, I. Mandel, P. Macias, A. Antoni, M. MacLeod and J. Law-Smith. We also thank the referee for very useful comments and suggestions. We thank the Niels Bohr Institute for its hospitality while part of this work was completed, and acknowledge the Kavli Foundation and the DNRF for supporting the 2017 Kavli Summer Program. E.R.-R. and R.W.E. thank the David and Lucile Packard Foundation, the Heising-Simons Foundation, and the Danish National Research Foundation (DNRF132) for support. R.W.E. is supported by the Eugene V. Cota-Robles Fellowship and National Science Foundation Graduate Research Fellowship Program (Award \#1339067). Any opinions, findings, and conclusions or recommendations expressed in this material are those of the authors and do not necessarily reflect the views of the National Science Foundation. The software used in this work was developed in part by the DOE NNSA ASC- and DOE Office of Science ASCR-supported Flash Center for Computational Science at the University of Chicago.} Resources supporting this work were provided  by the University of Copenhagen high-performance computing cluster funded by a grant from VILLUM FONDEN (project number 16599).

\bibliographystyle{aasjournal}
\bibliography{bib}
\end{document}